# Josephson junctions of Weyl semimetal $WTe_2$ induced by spontaneous nucleation of PdTe superconductor.

Manabu Ohtomo,[1]* Russell S. Deacon,[2,3] Masayuki Hosoda,[1,2] Naoki Fushimi,[1] Hirokazu Hosoi,[1] Michael D. Randle,[2] Mari Ohfuchi,[1] Kenichi Kawaguchi,[1] Koji Ishibashi,[2,3] and Shintaro Sato[1]

[1]*Fujitsu Research, Fujitsu Ltd., Atsugi, Kanagawa 243-0197, Japan.*

[2]*Advanced Device Laboratory and* [3]*Center for Emergent Matter Science (CEMS), RIKEN, Wako, Saitama 351-0198, Japan*

*E-mail: ohtomo.manabu@fujitsu.com

We report on the fabrication of Josephson junction devices with weak links utilizing the Weyl and higher-order topological semimetal $WTe_2$. We show that $WTe_2$/Pd contact annealed at a low temperature of 80°C did not exhibit superconducting properties because neither $WTe_2$ nor Pd are superconductors in the ground state. Upon 180°C annealing, spontaneous formation of superconducting PdTe due to Pd diffusion enabled us to obtain the interface between $WTe_2$ and superconductor suitable for the Josephson junction. This result is a facile technique to make a Josephson junction and induce Cooper pairs into topological telluride semimetals.

The Josephson junction (JJ) composed of a topological insulator (TI) is intensively studied as a platform to demonstrate topological superconductivity, which is one of the possible routes to realize Majorana particles[1-3] for future fault-tolerant Majorana qubits.[4-7] As for the topological superconductors, while nanowire-based approaches with magnetic field are advanced,[9,10] another approach using one dimensional conducting states protected by time-reversal symmetry in topological insulators (TI) or higher order topological insulators (HOTI) is becoming an alternative option for hosting Majorana particles.[11] The advantage of the TI approach is that they are expected to host single-mode conduction and exhibit insensitivity to disorder, as well as no requirement for applied magnetic fields. Among various materials proposed for TI and HOTI, the transition metal dichalcogenide $WTe_2$ has the unique property that it behaves as a two-dimensional TI in its single-layer 1$T$' phase[12] and a type-II Weyl semimetal in its bulk $T_d$ phase. The latter is also claimed to be a HOTI with characteristic hinge states.[13,14]

The recipe for a topological superconductor is to induce superconductivity into a TI or HOTI using the proximity effect from an s-wave superconductor in contact with the TI or HOTI.[11,15] Making good contact to the edge state of the TI or hinge state of the HOTI using superconducting electrodes is particularly important because any interface barrier would hamper the proximity effect and make the induced superconducting gap "soft",[16] which would invalidate topological protection.[17,18] This is problematic because the surfaces of s-wave superconducting materials are usually not stable in air. The surfaces of widely used Al or Nb, for example, are easily covered by a passive oxide film if a bottom contact geometry (electrodes are formed first before the lamination of $WTe_2$ on top) is employed. A top-contact geometry (electrodes are formed on top of $WTe_2$ crystals), on the other hand, is also challenging because of the surface oxidation of $WTe_2$. It is reported that the surface of $WTe_2$ is oxidized under ambient conditions, forming a surface oxide layer with a thickness of ~2.5 nm.[19] During the fabrication process of the top-contact electrodes, any exposure of the $WTe_2$ surface to oxygen or water should be avoided, which is not easy to achieve in a conventional nano-patterning process. A JJ device is also an ideal platform to study the contact problem such as the interface oxide layer. The observation of JJ characteristics is clear evidence of the proximity effect.

In this letter, we report on the facile fabrication of a good $WTe_2$-superconductor junction suitable for a topological JJ using spontaneous diffusion of Pd atoms at the $WTe_2$/Pd interface. A superconducting region is achieved at the $WTe_2$/Pd interface, even though neither $WTe_2$ nor Pd are superconducting in the ground state. From our Scanning

Transmission Electron Microscope (STEM) and X-ray photoemission spectroscopy (XPS) analysis, we conclude that a PdTe superconducting layer is formed at the $WTe_2$/Pd interface due to Pd diffusion at annealing temperatures as low as 180 °C.

A $T_d$-$WTe_2$ single crystal (purchased from HQ graphene, see Fig. 1a for the crystal structure) was cleaved using the standard adhesive tape method.[20] The cleaved crystal was transferred using the so-called Polydimethylsiloxane (PDMS) stamp method (Fig. 1b).[21] A polycarbonate (PC) adhesion layer was placed onto a dome of PDMS [Fig. 1b(i)] on a glass slide. This PC film is brought into contact with the target crystal and heated at 110°C for pick-up [Fig. 1b(ii)]. The crystal was then transferred to the target position on the device substrate and annealed at 180°C, releasing the PC from the PDMS. JJ devices were fabricated in the bottom contact geometry. $WTe_2$ was transferred onto pre-patterned Pd electrodes, fabricated using standard e-beam lithography (EBL) followed by electron beam deposition and lift-off. We took care not to contaminate the Pd electrode surface with resist, and the Pd electrode was formed as the last step of the substrate fabrication process. The melted PC polymer residue was intentionally left on the crystal as a passivation film in the bottom contact geometry [Fig. 1b(iii)]. The crystal cleavage and transfer processes were performed in a glove box filled with pure Ar (dew point less than -80°C). Two devices (device I and II) were prepared with PC/PDMS domes using a maximum annealing temperature ($T_a$) of 180°C. For device III, fabricated with low temperature annealing ($T_a$ = 80°C), the $WTe_2$ was directly exfoliated onto a PDMS stamp, which was then brought into contact with the prepatterned Pd contacts. The sample was heated to 80°C for one minute before slow delamination of PDMS at 30°C, leaving the flake in contact. Devices with the top-contact geometry were used to investigate annealing temperature dependence of the resistance. Top electrodes were fabricated by EBL and lift-off after the PC film residue on $WTe_2$ was removed using chloroform. The details of the sample fabrication techniques can be found in the supplementary information (SI).

The optical microscope image of the bottom contact device (device I) fabricated using PC stamp with 180°C annealing is shown in Fig. 1c. The channel length ($L$) and width ($W$) of the devices I and II are: $L$ = 1.47 μm, $W$ = 1.65 μm (device I), and $L$ = 1.50 μm, $W$ = 2.29 μm (device II). The other details of the device geometries are summarized in the SI. The temperature dependence of the device resistance (Fig. 1d) shows that the device with 180°C annealing (device II fabricated by PC/PDMS transfer) works as a JJ with $T_C$ = ~1.5 K, which reproduces the results of previous reports.[13,22] The $I$-$V$ relation [Fig. 1e(i)] extrapolates to zero, making it difficult to estimate the excess current. This is an unavoidable and well-established problem[23] when using narrow electrodes (~300 nm) to measure the Fraunhofer pattern without flux jumps. A Fraunhofer-like pattern is observed in both devices (device I and II) fabricated with 180°C annealing (plotted in Fig. 1e and SI), which indicates the interference of supercurrent in the JJ. Deviation from the standard Fraunhofer pattern (see equation S1 in the SI for the description) is observed in our samples, which is often attributed to an inhomogeneous supercurrent distribution in the channel, which may include hinge states. Discussing the appearance of hinge states solely from the interference pattern is often controversial[24] and beyond the scope of this study. These results are in clear contrast with the 80°C-annealed device (device III; fabricated by PDMS transfer), which does not demonstrate any feature consistent with a superconducting transition (Fig. 1d). After annealing device III at 200°C in an Ar atmosphere, we confirmed a superconducting transition. This result indicates that 180-200°C annealing plays a key role in making the $WTe_2$/Pd contact superconducting.

Cross-sectional STEM images of $WTe_2$/Pd interfaces in device II, fabricated using PC stamp with 180°C annealing, is shown in Fig. 2. There are several notable changes in the STEM images such as swelling of the $WTe_2$ and roughening of the Pd electrode surfaces [Fig. 2a(ii-iii)]. While the typical surface roughness (arithmetical mean deviation $R_a$) of the Pd electrodes as deposited was small and in the order of ~0.38 nm, considerable roughening of the Pd surface was observed after contacting with $WTe_2$ and annealing. Similar swelling was observed in all the electrodes (denoted A-E) in this device (Fig. 2a). The energy dispersive X-ray spectroscopy (EDX) profile shown in Fig. 2b indicates that Pd diffusion triggered the $WTe_2$ swelling. While there are some $WTe_2$ layers without Pd diffusion, Pd diffusion was observed in the interface region of $WTe_2$ [see Fig. 2b(iii) for the magnified image] as well as in the swollen $WTe_2$. Line profiles from STEM-EDX are shown in Fig. 2c. The crystals were separated into two layers, one assigned to $WTe_2$ as determined from the EDX profile. The atomic distribution of Pd shown in the scans indicates that the other layer mainly consists of Pd and Te, while a density gradient of W is also observed. It is noteworthy that the EDX signals of oxygen at the $WTe_2$/PdTe interface were negligible, although the $WTe_2$ surface is normally easily oxidized on its own. The Pd signal is also observed on the $WTe_2$ surface, which is not in contact with Pd [Fig. 2b(iii) and Fig. 2c], while the $WTe_2$ bulk with layered structure mainly consists of W and Te. The origin of Pd on the $WTe_2$ surface is probably side-wall diffusion of the $WTe_2$ crystal. This result implies a high degree of Pd diffusion in $WTe_2$ crystals.

Magnified STEM images of electrode C are shown in figures 2a(ii) and 2d(i-iii). The corresponding FFT analysis of the STEM images in Fig. 2d(i-iii) are shown in Fig. 2d(iv-vi), respectively. The initial stage of the Pd diffusion appears to be Pd intercalation because swelling starts in the middle of the $WTe_2$ layers [Fig. 2d(i)]. While some of the swollen area consisting of Pd and Te were amorphous, the FFT analysis of the STEM images reveals the areas with crystalline structure other than $WTe_2$ or Pd in the middle of the interfaces [Fig. 2d(ii)]. From the FFT analysis [Fig. 2d(v)] and electron diffraction simulation [Figure S8 in the SI], we determine that these polycrystals are either PdTe or $PdTe_2$, which have hexagonal (space group $P6_3/mmc$) and trigonal (space group $P\overline{3}m1$) crystal structure, respectively, with almost identical lattice constants along the $a$ and $b$ axes [See Table S2 and Figure S5 in the SI for the structures]. Both PdTe and $PdTe_2$ are superconductors with $T_C$ of 2.3-4.5 K[25,26] and 2.0 K,[27] respectively. A similar structure was observed below $WTe_2$ in electrode D (Fig. 2e), in which the FFT pattern is assigned to PdTe from systematic absences. Explicitly, the $(000\overline{1})$ and $(000\overline{2})$ diffractions are missing in Fig. 2e(ii) (indicated by broken line arrows), which is evidence for PdTe [See Figure S8 in the SI for the simulation]. This is in contrast with another telluride-TI material, $(Bi_xSb_{1-x})_2Te_3$ (BST), where intermixing of Pd and Te results in the self-formation of superconducting $PdTe_2$.[23,28] In all electrodes, we confirmed that $WTe_2$ without Pd diffusion maintained its $T_d$ structure as shown in Fig. 2e(iii).

The formation of PdTe is further confirmed by XPS (Fig. 3a). The sample prepared for XPS analysis consists of cleaved $WTe_2$ crystals on $SiO_2$ with a thin Pd film (~5 nm) deposited on top (See SI for the other details). Pt $4d_{3/2}$ peaks in the spectra were derived from the alignment markers on the sample. Three spectra are shown in Fig. 3a, which were taken (i) as deposited, after (ii) 100°C and (iii) 180°C annealing, respectively. Sample annealing was performed *ex-situ* in a glove box with Ar atmosphere. The position of Pd $3d_{3/2}$ and $3d_{5/2}$ peaks were observed to shift as the sample was annealed. Table 1 shows the binding energies (BE) observed experimentally, and BEs calculated by density functional theory (DFT)[29] using the DFT package OpenMX[30] with pseudoatomic orbital assuming core holes in Pd (see the SI for the details). For metals, the absolute binding energy ($E_b{}^{metal}$) is estimated using the following formula:

$$E_b{}^{metal} = E_f^{(0)}(N) - E_i^{(0)}(N) \quad (1)$$

where $E_i^{(0)}(N)$ and $E_f^{(0)}(N)$ are the intrinsic total energy of the ground state and the excited state with $N$ electrons, respectively. The calculated BEs for PdTe and $PdTe_2$ (Table 1) reproduce the experimentally determined BEs reported from previous reports on Pd bulk and $PdTe_2$[8] (see Fig. S5 in the SI for the models). After 180°C annealing, the BEs of Pd $3d_{5/2}$ and $3d_{3/2}$ were shifted to the higher BE side (335. 8 eV and 341.1 eV, respectively), close to the calculated BE of PdTe (335.56 eV and 341.17 eV for $3d_{5/2}$ and $3d_{5/2}$, respectively). These XPS results clearly indicate that the PdTe is formed at the $WTe_2$/Pd interfaces during 180°C annealing, and it is consistent with our STEM analysis.

The effect of Pd diffusion was also confirmed in the two-probe resistance measurement of the top-contact devices (Fig. 3b). The top-contact devices were fabricated by depositing 50 nm thick Pd on the $WTe_2$ crystals. The device was measured in vacuum (~$10^{-5}$ Pa) at 300 K and annealing was performed *in-situ*. The dependence of the two-probe resistance ($R_{2\text{probe}}$) normalized by channel widths ($W$) against annealing temperature ($T_{\text{anneal}}$) is plotted in Fig. 3b at various $L$. Contact resistance ($R_C$) was evaluated by the transfer length method (TLM) and is also shown in Fig. 3b. $R_{2\text{probe}}$ and $R_C$ monotonically decreased as annealing temperature increased, with a particularly large decrease after annealing at 100°C. The decrease in both $R_{2\text{probe}}$ and $R_C$ was almost saturated at 150°C, yielding $R_C W$ ~18 Ωμm with a ~61.6 nm thick crystal. These results suggest that Pd diffusion and PdTe crystallization took place after annealing at 100-150°C. It should be noted that among the junctions measured, $R_{2\text{probe}}$ of seven junctions decreased, while that of three junctions on the same crystal slightly increased after annealing at 423 K. We assume that this is associated with the formation of amorphous $PdTe_x$ at the interface, which was observed in the STEM image of two out of six electrodes (D and E) in the bottom contact device [See Figure S9 and S13 in the SI]. The mechanism for amorphization and how to control it are not yet fully understood.

The experimental results presented above indicate the formation of a superconducting PdTe polycrystal layer at the $WTe_2$/Pd interface upon 180°C annealing in our samples. Let us briefly consider the charge transfer, which is one of the possible mechanisms for $WTe_2$/Pd interfacial superconductivity suggested in the literature.[13,22)] Our DFT calculations reveal considerable charge transfer mainly due to Pd chemisorption on $WTe_2$ (See Fig. S7 and Table S3 in the SI for the DFT results) when a clean and abrupt $WTe_2$/Pd interface is assumed. The amount of charge transfer (Table S3 in the SI) is sufficient to induce superconductivity in monolayer $1T'$-$WTe_2$,[31)] but it is not evident that it is enough to induce superconductivity in the bulk material. This calls for a quantitative investigation into the phase diagram of these materials with respect to changes in charge doping, which is a subject of future work. From our experimental results, we conclude that the formation of PdTe is a plausible and straightforward origin of the observed superconductivity at $WTe_2$/Pd interfaces.

In summary, we report on the spontaneous nucleation of superconducting PdTe at $WTe_2$/Pd interfaces, which can be utilized to create Josephson junction devices of the Weyl semimetal $WTe_2$. Pd diffusion during 180°C annealing was confirmed by STEM and EDX analysis, while the formation of PdTe was evidenced by electron diffraction and XPS including first-principles estimation of binding energies. The STEM analysis revealed that the oxide layer at the $WTe_2$/PdTe interface was negligible, which implies a clean interface between superconductors and HOTIs. This method is a novel technique to make good superconducting contacts to topological telluride materials for topological Josephson junction devices.

**Acknowledgments**

A part of this work was conducted at NIMS Nanofabrication Platform, supported by "Nanotechnology Platform Program" of the Ministry of Education, Culture, Sports, Science and Technology (MEXT), Japan, Grant Number 21A019. This work was supported by JSPS Grant-in-Aid for Scientific Research (A) (No. 19H00867), Grant-in-Aid for Scientific Research (B) (No. 19H02548) and Grants-in-Aid for Scientific Research (S) (No. 19H05610). We thank Ms. Yumiko Shimizu (Toray Research Center, Inc.) for the STEM measurement.

## Figure Captions

**Fig. 1.** (a) The crystal structure of $T_d$-$WTe_2$, where the unit cell is encapsulated by the blue sold lines. (b) Schematic diagram of the device fabrication process. (c) Optical microscope image of the $WTe_2$/Pd bottom contact device fabricated by PC stamp method (device I). (d) Zero-bias resistance ($R$) as a function of temperature ($T$) in $H$ = 0 at various maximum annealing temperature during the process ($T_a$). (e) (i) $I$-$V$ characteristics of device III after 200°C annealing. The measurement is performed at 250 mK in a three-probe configuration. The series resistance from the contacts (~240 Ω) is subtracted from the data. (ii) Two-dimensional (2D) mapping of the differential resistance ($dV/dI$) of device I plotted in the $I$ vs $H$ plane (measured at 70 mK). The remanent magnetic field of the superconducting magnet ($H_0$ = 11.8 mT) is subtracted.

**Fig. 2.** (a) (i) The cross-section STEM image of device II ($T_a$ = 180°C). Schematic diagram of the device is also shown. The magnified image of electrode C and D is shown in (ii) and (iii), respectively. The scale bars in (ii) and (iii) are 100 nm. (b) STEM-EDX mapping of electrode B for (i) Te, (ii,iii) Pd and (iv) W. The area indicated by white square is magnified in (iii). (c) EDX line-scan profile of the interface. The line scan is corrected from the area indicated by yellow lines in the STEM image (lower panel), which was taken at the position indicated by white broken line in Fig. 2b(iii). (d) High-resolution HAADF-STEM (i) and BF-STEM (ii, iii) images of electrode C. See Fig. 2a(ii) for the position. The FFT pattern obtained from the area indicated by a yellow circle in (i-iii) is shown in (iv-vi), respectively. (e) (i) High-resolution BF-STEM image of electrode D. See Fig. 2a(iii) for the position. The FFT pattern obtained from the area indicated by a yellow circle in (i) is shown in (ii, iii). The missing diffraction is indicated by a broken line arrow in (ii).

**Fig. 3.** (a) The Pd 3*d* XPS spectra of a thin Pd film on a $WTe_2$ flake (i) as deposited, after annealing in Ar atmosphere at (ii) 100°C and (iii) 180°C. The Pt 4*d* peak is derived from alignment markers on the sample. (b) Annealing temperature ($T_{anneal}$) dependence of two probe resistance ($R_{2probe}$) and contact resistance ($R_C$) in the top-contact device. The $R_{2probe}$ and $R_C$ are normalized by channel width $W$.

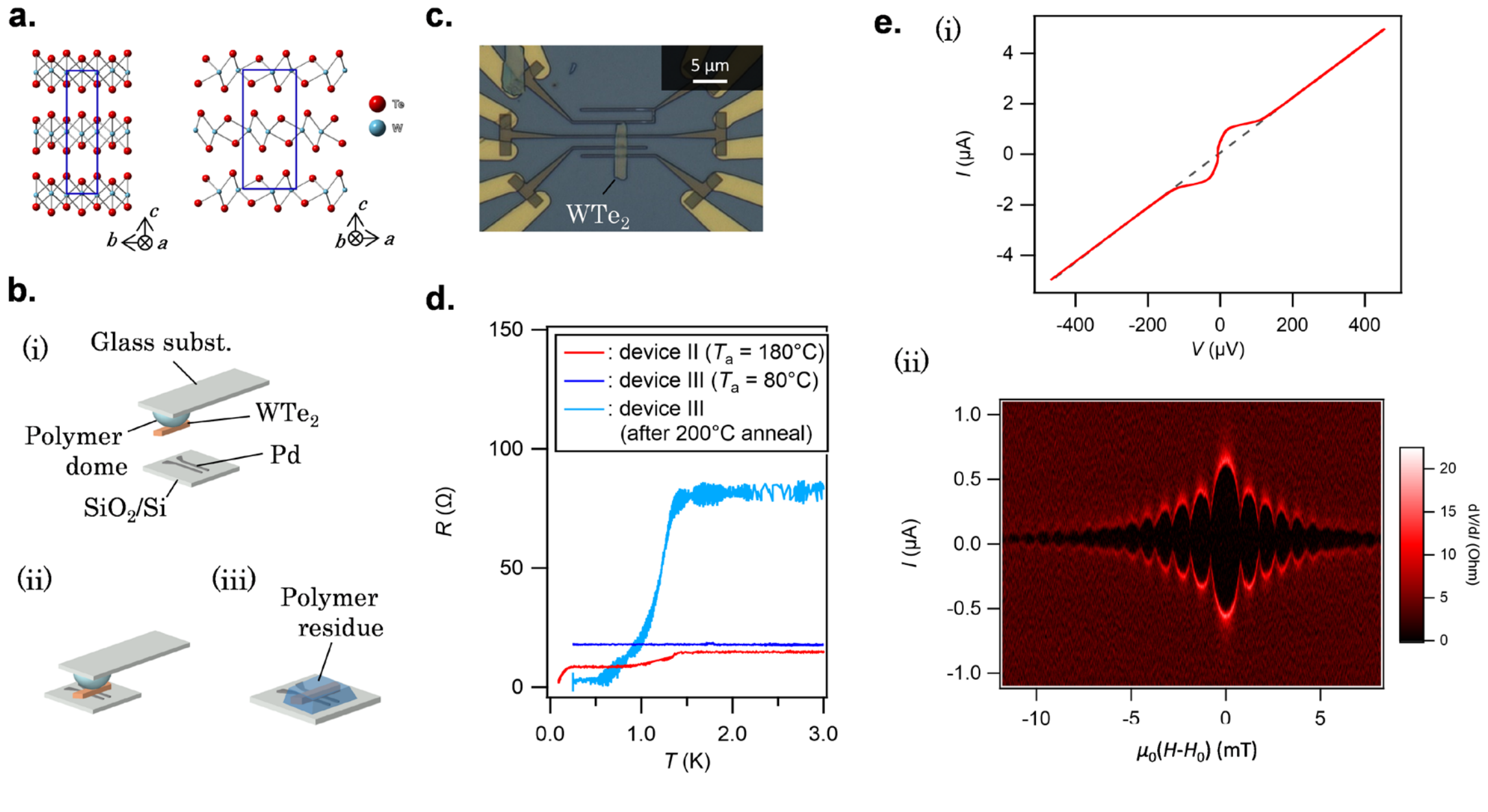

Fig. 1

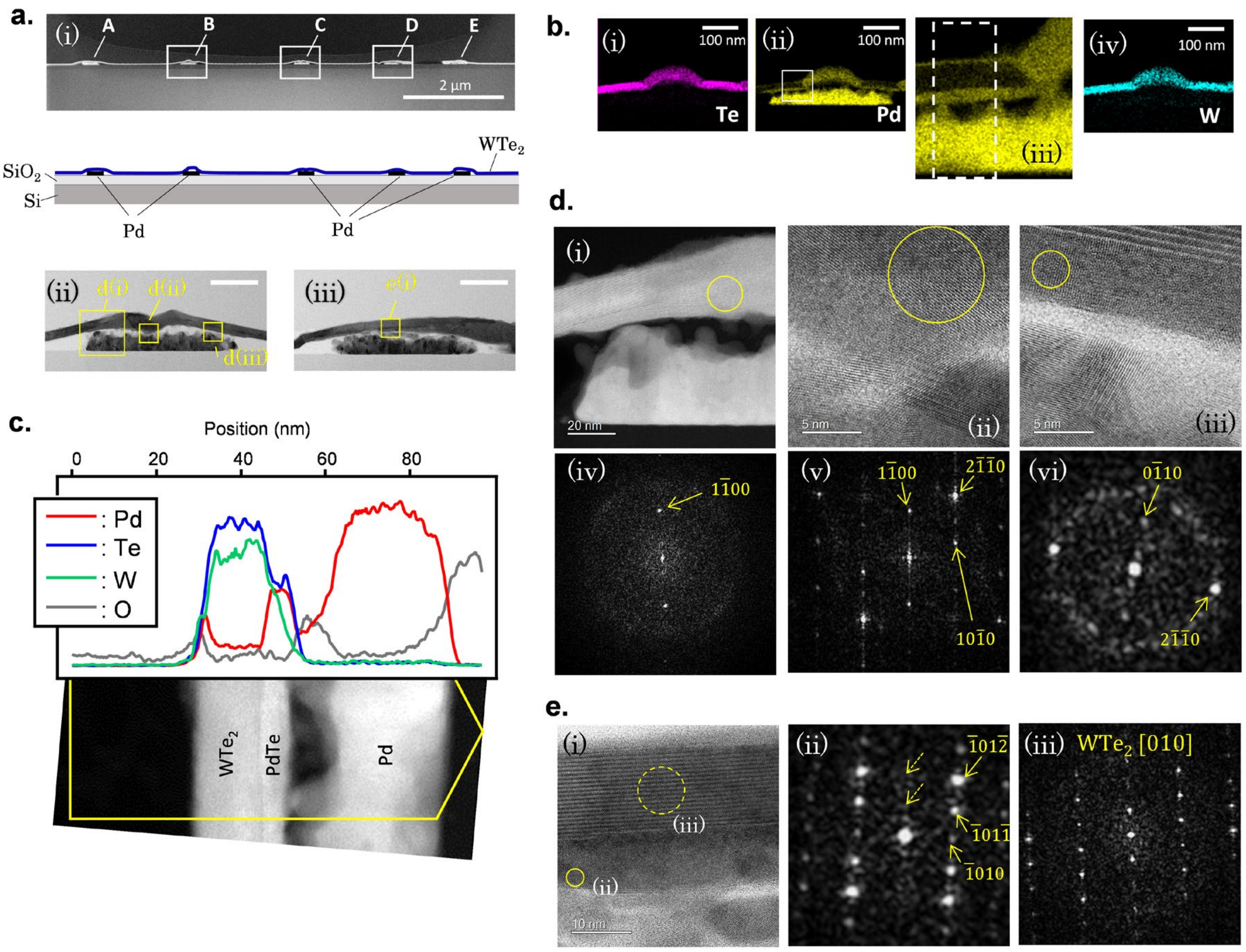

Fig. 2

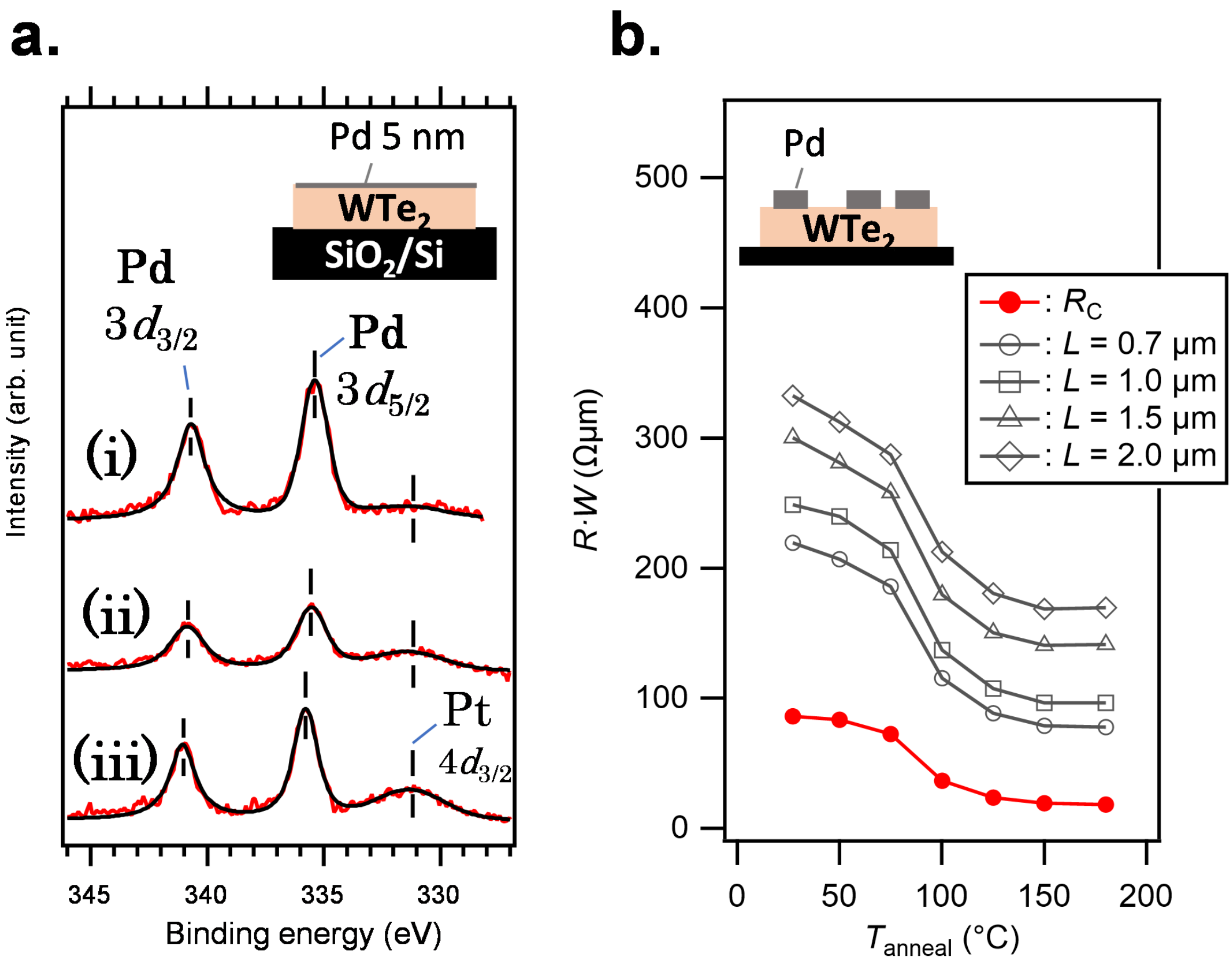

Fig. 3

**Table I.** The binding energy (B.E.) of Pd 3*d* doublet.

| | B.E. of Pd $3d_{5/2}$ (eV) | B.E. of Pd $3d_{3/2}$ (eV) |
|---|---|---|
| Pd bulk (calc.) | 335.15 | 340.56 |
| PdTe (calc.) | 335.56 | 341.17 |
| $PdTe_2$ (calc.) | 336.22 | 341.79 |
| $PdTe_2$ (exp.) [8] | 336.7 | 341.9 |
| as depo. (exp.) | 335.4 | 340.7 |
| 100°C anneal (exp.) | 335.5 | 340.9 |
| 180°C anneal (exp.) | 335.8 | 341.1 |

Supplementary information

# Josephson junctions of Weyl semimetal $WTe_2$ induced by spontaneous nucleation of PdTe superconductor.

Manabu Ohtomo,[1]* Russell S. Deacon,[2,3] Masayuki Hosoda,[1,2] Naoki Fushimi,[1] Hirokazu Hosoi,[1] Michael D. Randle,[2] Mari Ohfuchi,[1] Kenichi Kawaguchi,[1] Koji Ishibashi,[2,3] and Shintaro Sato[1]

[1]*Fujitsu Research, Fujitsu Ltd., Atsugi, Kanagawa 243-0197, Japan.*

[2]*Advanced Device Laboratory and* [3]*Center for Emergent Matter Science (CEMS), RIKEN, Wako, Saitama 351-0198, Japan*

# 1. Materials and Methods

## 1.1. Device fabrication procedure and Fraunhofer pattern

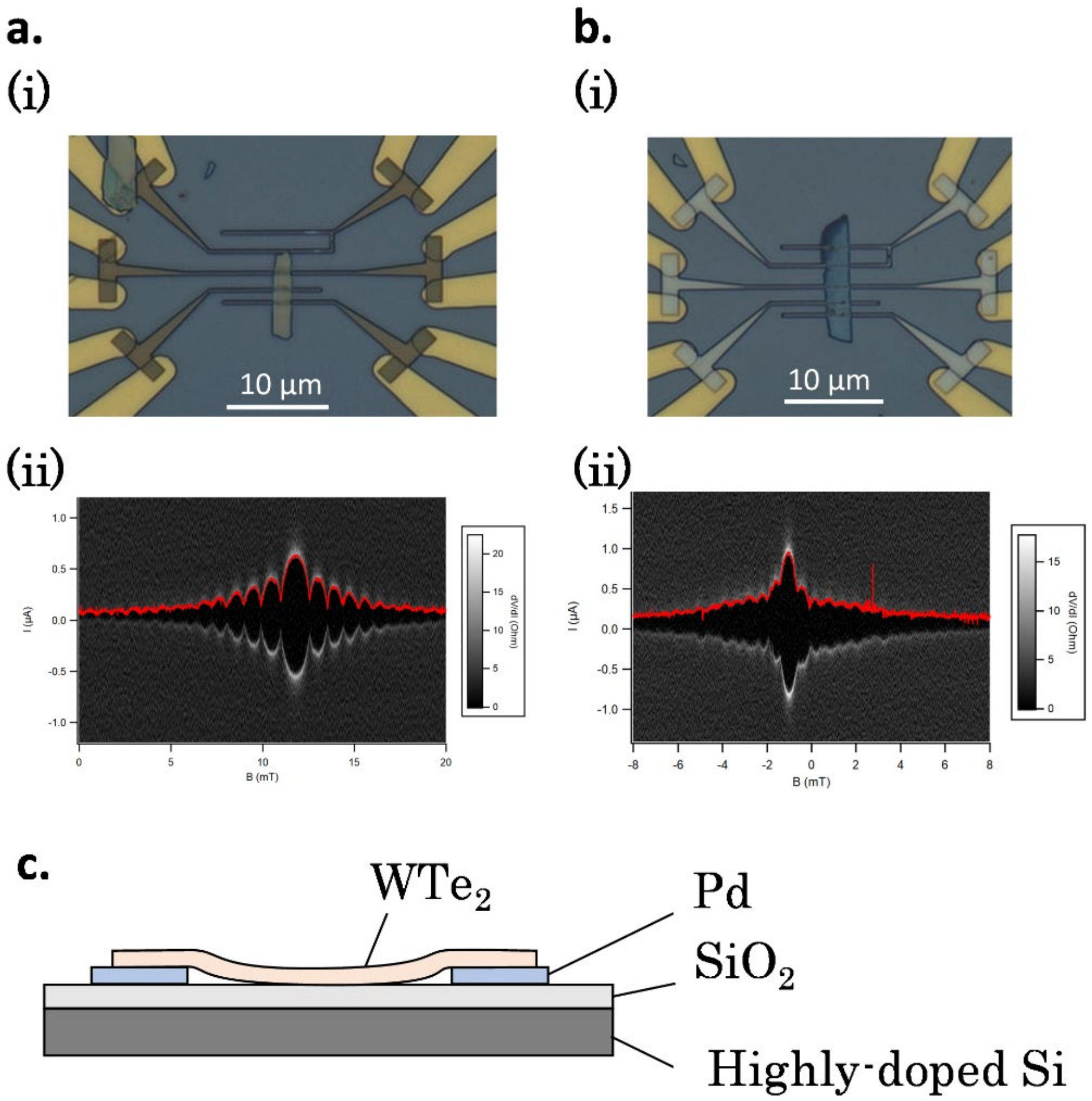

**Figure S1 (i) The optical microscope image and (ii) Fraunhofer pattern for (a) device I and (b) device II. Switching current is shown in red solid lines. (c) Schematic drawing of the device structure (bottom contact).**

The fabrication process of bottom contact devices is summarized as follows. A substrate is highly doped Si with thermal oxide $SiO_2$ thickness of 285 nm. Probing pads and lead lines consisting of Au/Ti double layer were patterned on $SiO_2$ by photolithography. PMMA A4 resist was then spin-coated on the substrate followed by pre-baking at 180°C. After electron beam exposure, developing was performed by mixture of MIBK (methyl isobutylketone) and IPA (isopropyl alcohol). Pd layer (~30 nm thick) was deposited and lift off was performed by acetone. The channel length ($L$), width ($W$) and crystal thickness ($T$) of the device I-III are: $L$ = 1.47 μm, $W$ = 1.65 μm, $T$ = 52 nm (device I), $L = 1.50$ μm, $W = 2.29$ μm, $T = 18.3$ nm (device II), and $L = 468$ nm, $W = 4.45$ μm, $T = 12.5$ nm (device III).

Optical microscope images and Fraunhofer-like patterns of the device I and II are shown in Figure S1. The standard Fraunhofer pattern is usually described as:

$$I_C(H) = I_{C0}\left|sin\left(\pi\,\Phi_J/\Phi_0\right)/\left(\pi\,\Phi_J/\Phi_0\right)\right| \qquad \text{(S1)}$$

where $I_{C0}$, $\Phi_J$, and $\Phi_0$ are the critical current at $H = 0$, magnetic flux penetration into the junction, and flux quantum $\Phi_0 = h/2e$, respectively. As mentioned in the main text, a deviation of the interference pattern from the standard Fraunhofer pattern [equation (S1)] is often regarded as a manifestation of an inhomogeneous supercurrent distribution such as hinge states. We note that determination of the supercurrent distribution solely from the interference pattern from analysis of the FFT in this manner is controversial,[1] and we consider a conclusion based upon it beyond the scope of this study.

### 1.2. XPS measurement

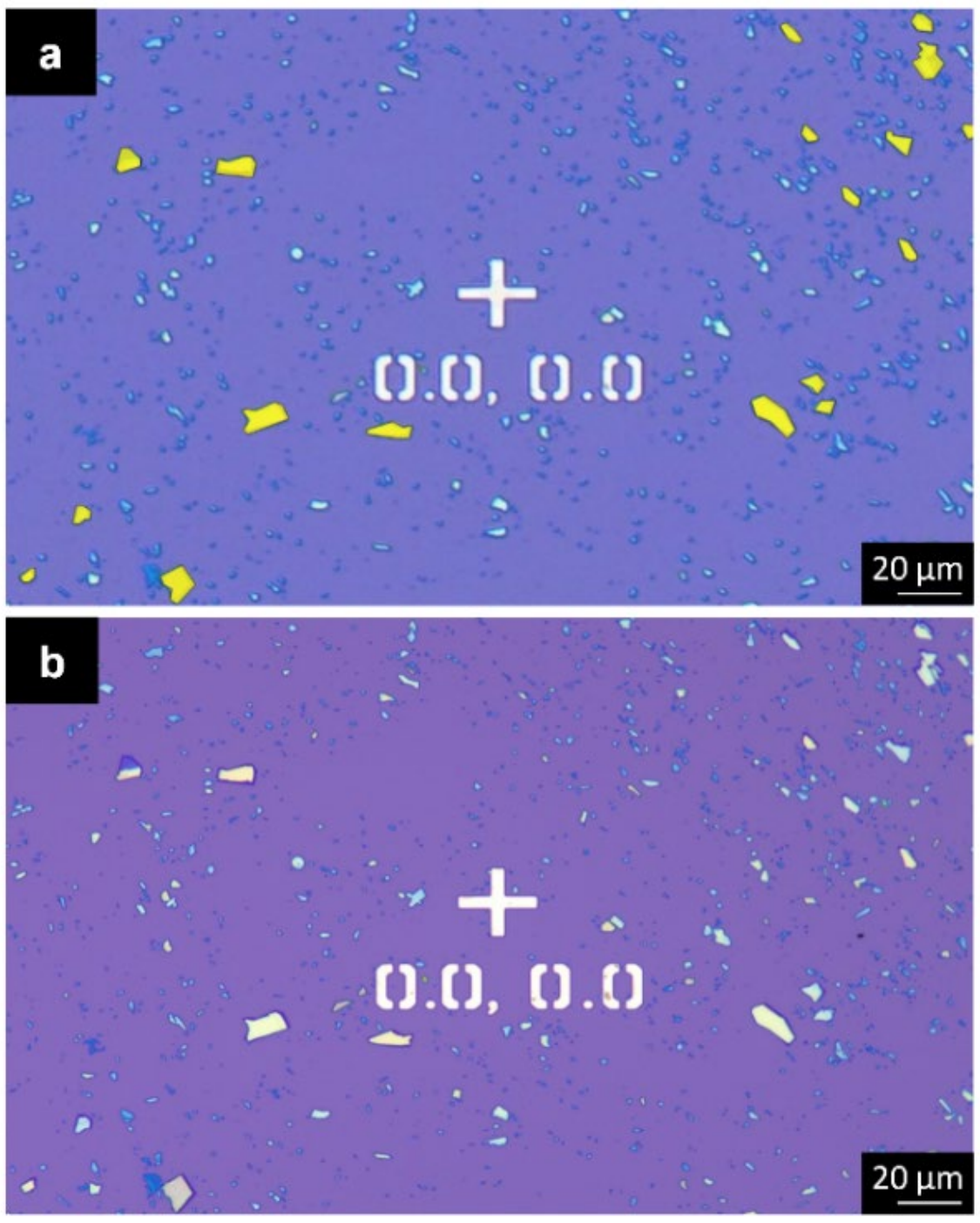

**Figure S2 Optical microscope image of the XPS sample. Thin Pd film is deposited on relatively large crystals filled by yellow polygon overlay in (a). An image of the same area after Pd deposition is shown in (b). The alignment marker is shown in the center of the images.**

X-ray photoemission spectroscopy (XPS) experiments (Kratos AXIS-HSi) are performed at $2 \times 10^{-7}$ Pa at room temperature using monochromatized Al Kα line ($h\nu$ = 1486.6 eV) radiation. For XPS measurements, annealing was performed in a glove box under Ar atmosphere. The sample was transferred in air from the glove box to an XPS chamber. Energy calibration was performed using the O 1s, C 1s and Fermi edges in each sample. The energy resolution was set to 550 meV for all spectra and background subtraction was performed using the Shirley method.

The details of the XPS samples are summarized as follows (Figure S2). The $WTe_2$ single crystal is cleaved by the standard tape method and then laminated onto a $SiO_2$/Si substrate in an Ar glove box. The $SiO_2$/Si substrate has prepatterned alignment markers consisting of 50 nm thick Pt film (shown in the middle of the images in Figure S2), which is detected in our spectra. The Pd layer is then patterned using electron beam lithography (EBL) with PMMA resist only on the large $WTe_2$ flakes (indicated by yellow overlay in Figure S2a). The 5 nm thick Pd film is then prepared by electron beam evaporation and lift off. It should be noted that not all the $WTe_2$ flakes are covered by Pd, i.e., Te and W spectra include signals from oxidated $WTe_2$ flakes without Pd passivation.

### 1.3. The outline of TLM Device

The details of the TLM device are summarized as follows. The top-contact device is fabricated after $WTe_2$ transfer using a PC/PDMS polymer dome. The PC residue is removed by chloroform before spin coating in PMMA resist. The electrode patterning is performed by standard EBL and lift-off process. The electrode consists of 50 nm thick Pd film. The optical microscope image and AFM image are shown in Figure S3a (i) and (ii), respectively.

The cross-section STEM images of the TLM device after 180°C annealing are shown in Figure S3c. The $WTe_2$ crystal in part not in contact with Pd maintained its characteristic layered structure, while the layered structure was lost in the region (indicated by white arrow in Figure S3c) in contact with Pd. This result indicates that the decrease in both $R_{2probe}$ and $R_C$ shown in Fig. 3b was induced by Pd diffusion during annealing.

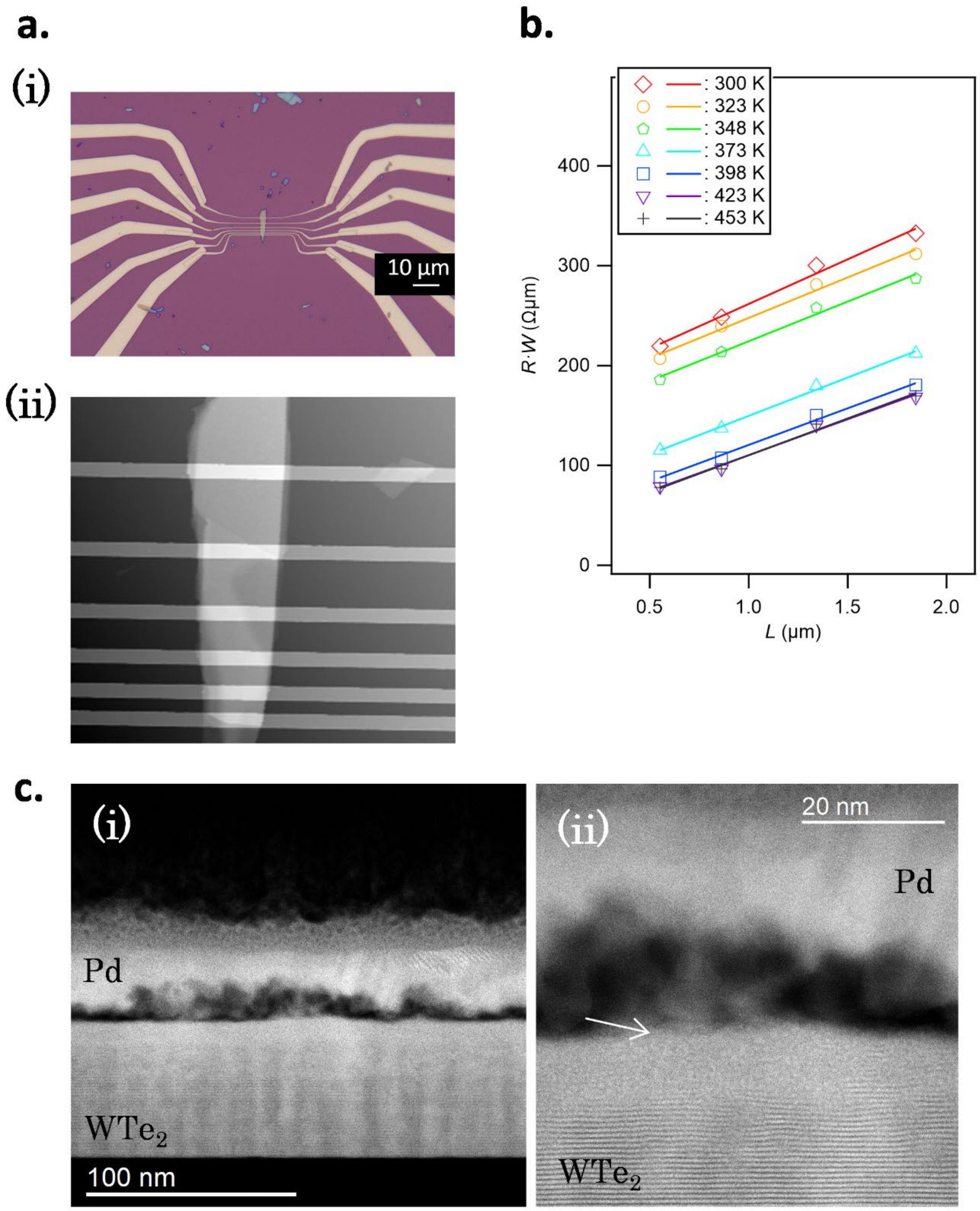

Figure S3 (a) Optical microscope image (i) and AFM image (ii) of the TLM device. The thickness of the crystal is ~60 nm and the scanning area of AFM image is 10μm×10μm. (b) The product of two-probe resistance (*R*) and channel width (*W*) plotted against channel length (*L*) after various annealing temperature. (c) Cross-section STEM image of the TLM device after 180°C annealing. The magnified image of (i) is shown in (ii).

## 2. First-principles calculations

### 2.1. Computational Details.

All computational calculations were performed using the density functional theory (DFT) code OpenMX.[2] The exchange−correlation potential was treated within the generalized gradient approximation using Perdew−Burke−Ernzerhof functional (GGA-PDE).[3] Electron−ion interactions were described by norm-conserving pseudopotentials with partial core correction.[4,5] Pseudoatomic orbitals centered on atomic sites were used as the basis function set.[6] The basis functions used for structural optimization and electronic structure calculations are summarized in Table S1. The van der Waals corrections were included using a semiempirical DFT-D2 method.[7] The effective Screening Medium (ESM) method was used for handling the repeated slab model periodic in surface parallel directions.[8,9] Geometry optimizations were performed under a three-dimensional periodic boundary condition. Convergence criteria for forces acting on atoms were set to 0.01 eV/Å for geometry optimization.

**Table S1 A set of the pseudo atomic orbitals (PAO) basis functions**

| | **structural optimization** | **electronic structure calculations** |
|---|---|---|
| **W** | W7.0-s3p2d2 | W7.0-s3p2d2f1 |
| **Te** | Te7.0-s3p2d2 | Te7.0-s3p2d2f1 |
| **Pd** | Pd7.0-s3p2d1 | Pd7.0-s3p2d2 |

## 2.2. Optimized structure of $WTe_2$ adsorbed on Pd surface

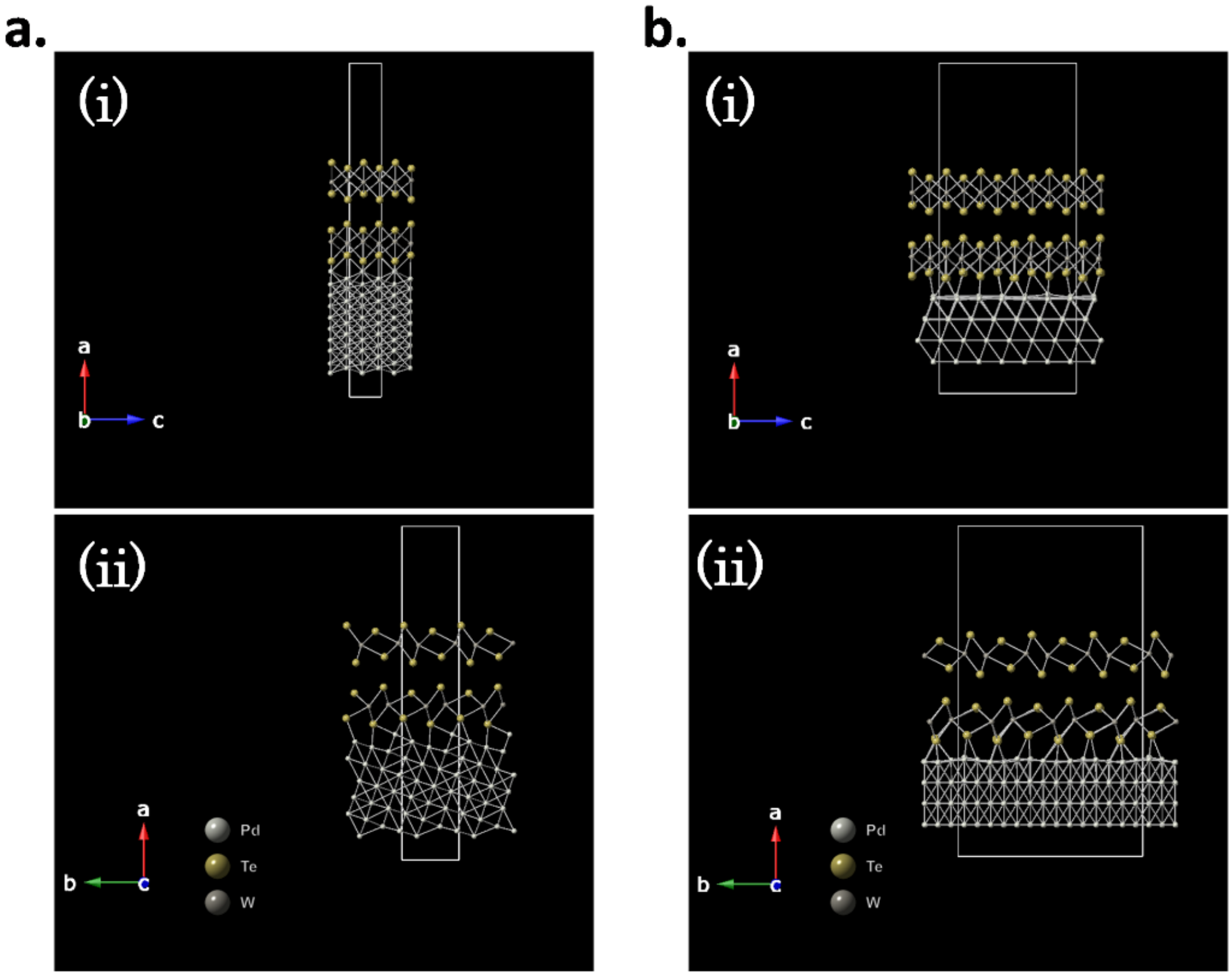

**Figure S4 The optimized structure for (a) (1×1)-$WTe_2$/Pd(1$\bar{2}\bar{3}$) and (b) (4×3)-$WTe_2$/Pd(111). The unit cell is indicated by white solid line.**

The optimized structures of $WTe_2$ adsorbed on Pd surfaces are summarized as follows. We prepared two models for $WTe_2$/Pd: (1×1)-$WTe_2$/Pd(1$\bar{2}\bar{3}$) and (4×3)-$WTe_2$/Pd(111). In both models, it is revealed that $WTe_2$ is chemisorbed on Pd. The Pd-Te bond length in both models is comparable or smaller than Pd-Te bond length in PdTe (2.86 Å) or $PdTe_2$ (2.76 Å). In Figure S4, PdTe bonds are depicted between Pd and Te atoms closer than 2.86 Å. The high reactivity of Pd to Te is probably the driving force of the Pd diffusion.

## 2.3. The details of the XPS binding energy estimation

The XPS binding energy (BE) estimation summarized in Table 1 was performed by Non-collinear DFT as follows. Three models are prepared for BE estimation: $WTe_2$/Pd($1\bar{2}\bar{3}$), PdTe, and $PdTe_2$. The structures of these models are optimized under a periodic boundary condition with the corresponding basis functions specified by Pd7.0_3d-s3p2d3f1 and Te7.0-s3p2d2f1. The calculation with core holes is performed using basis functions specified by Pd7.0_3d_CH-s3p2d3f1. Fully relativistic pseudopotentials are used with spin-orbit coupling. The models used for the XPS binding energy estimation are shown in Figure S4(a) and Figure S5. All models are optimized using a three-dimensional periodic boundary condition until the convergence criteria of 0.01 eV/Å is achieved on a *k* grid of 6×6×5. Lattice constants of PdTe and $PdTe_2$ are summarized in Table S2.

**Table S2 The optimized lattice constant for PdTe and $PdTe_2$.**

| | PdTe | $PdTe_2$ |
|---|---|---|
| $a = b$ | 4.1414 Å | 4.2700 Å |
| $c$ | 5.4781 Å | 5.7974 Å |
| $\alpha = \beta$ | 90.000° | 90.000° |
| $\gamma$ | 120.000° | 120.000° |

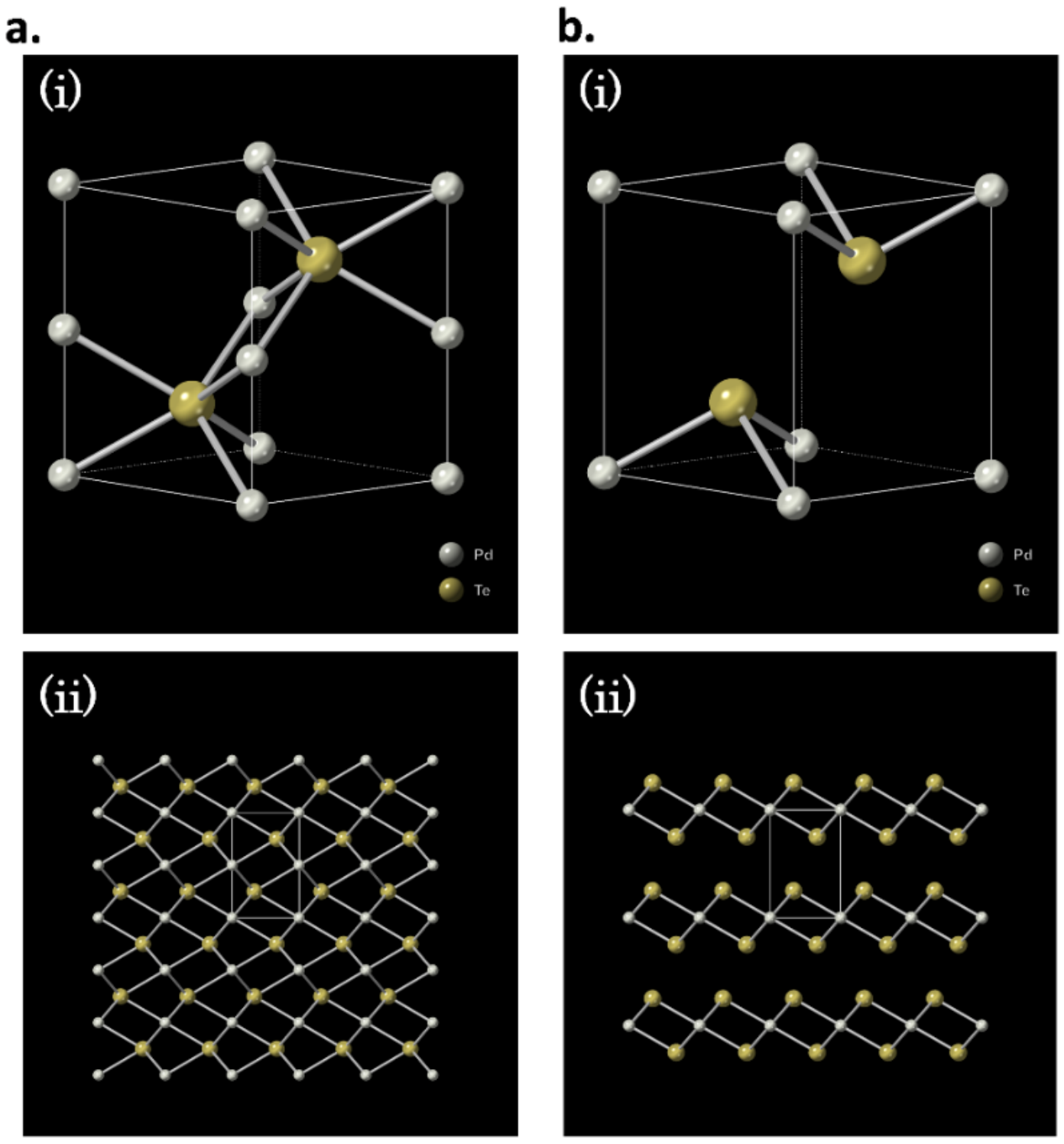

Figure S5. The crystal structure of PdTe (a) and $PdTe_2$ (b).

## 2.4. Bader charge analysis

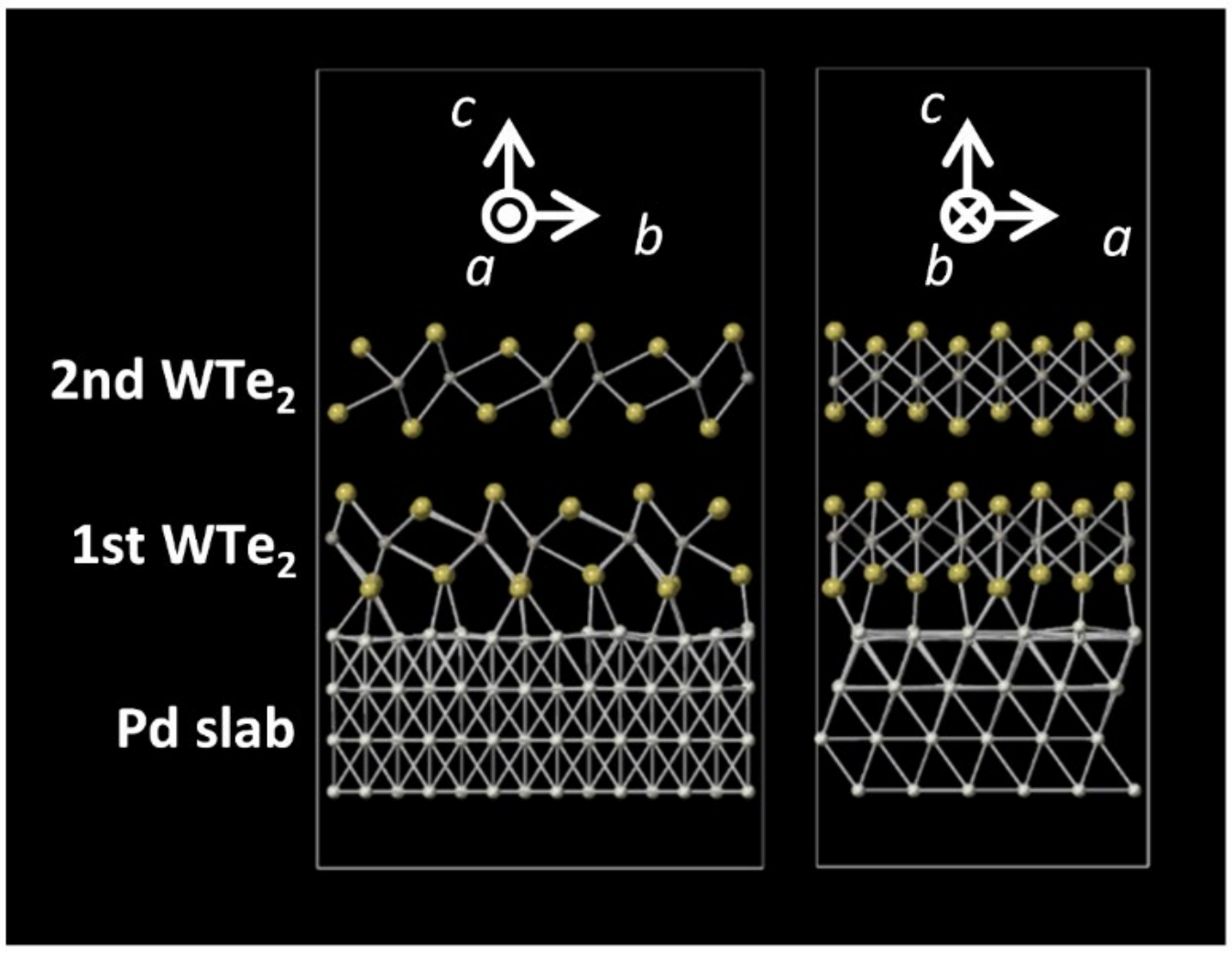

Figure S 6 The optimized structure of $WTe_2$/Pd(111) supercell.

In this section, the charge transfer in the $WTe_2$/Pd system is investigated using DFT. The model used is indicated in Figure S 7, in which $WTe_2$/Pd without Pd diffusion is considered. The purpose of this investigation is to evaluate charge transfer at the $WTe_2$/Pd interface, which could induce superconductivity. The Bader charge analysis[10,11] is applied to the electron density distribution data before and after $WTe_2$/Pd contact. The differential charge density induced in each layer of Pd and $WTe_2$ after making $WTe_2$/Pd contact is summarized in Table S3.

**Table S3 The differential electron density (unit: $cm^{-2}$) after making $WTe_2$/Pd contact.**

| | Pd slab | 1st $WTe_2$ | 2nd $WTe_2$ | 3rd $WTe_2$ | 4th $WTe_2$ |
|---|---|---|---|---|---|
| Bilayer $WTe_2$/Pd | $+1.09\times10^{14}$ | $-1.12\times10^{14}$ | $+3.52\times10^{12}$ | - | - |
| Quatrolayer $WTe_2$/Pd | $+1.09\times10^{14}$ | $-1.13\times10^{14}$ | $+3.68\times10^{12}$ | $+1.34\times10^{11}$ | $-2.24\times10^{10}$ |

## 3. Electron diffraction simulations

The electron diffraction simulation is performed using SingleCrystal software. The diffraction comparable to the FFT pattern of STEM images in the main text [Fig. 2d(iv-vi) and Fig. 2e(ii)] are summarized in this section. The simulation is performed using a crystal structure optimized by DFT as shown in Figure S5.

**Table S 4 The list of *d*(*hkl*) (in unit of nm) derived from experiment and simulation.**

| h | k | l | *d*(hkl) (exp.) | *d*(hkl) for $PdTe_2$ (sim.) | *d*(hkl) for PdTe (sim.) |
|---|---|---|---|---|---|
| 0 | -1 | 0 | 3.7 | 3.4884 | 3.5958 |
| 1 | -1 | 0 | 3.55 | 3.4884 | 3.5958 |
| -1 | 0 | 0 | 3.5 | 3.4883 | 3.5958 |
| 1 | 0 | 0 | 3.43 | 3.4883 | 3.5958 |
| -1 | 0 | -1 | 2.94 | 2.8825 | 3.0369 |
| -1 | 0 | -2 | 2.16 | 2.0633 | 2.2267 |
| 2 | -1 | 0 | 2.13 | 2.014 | 2.076 |

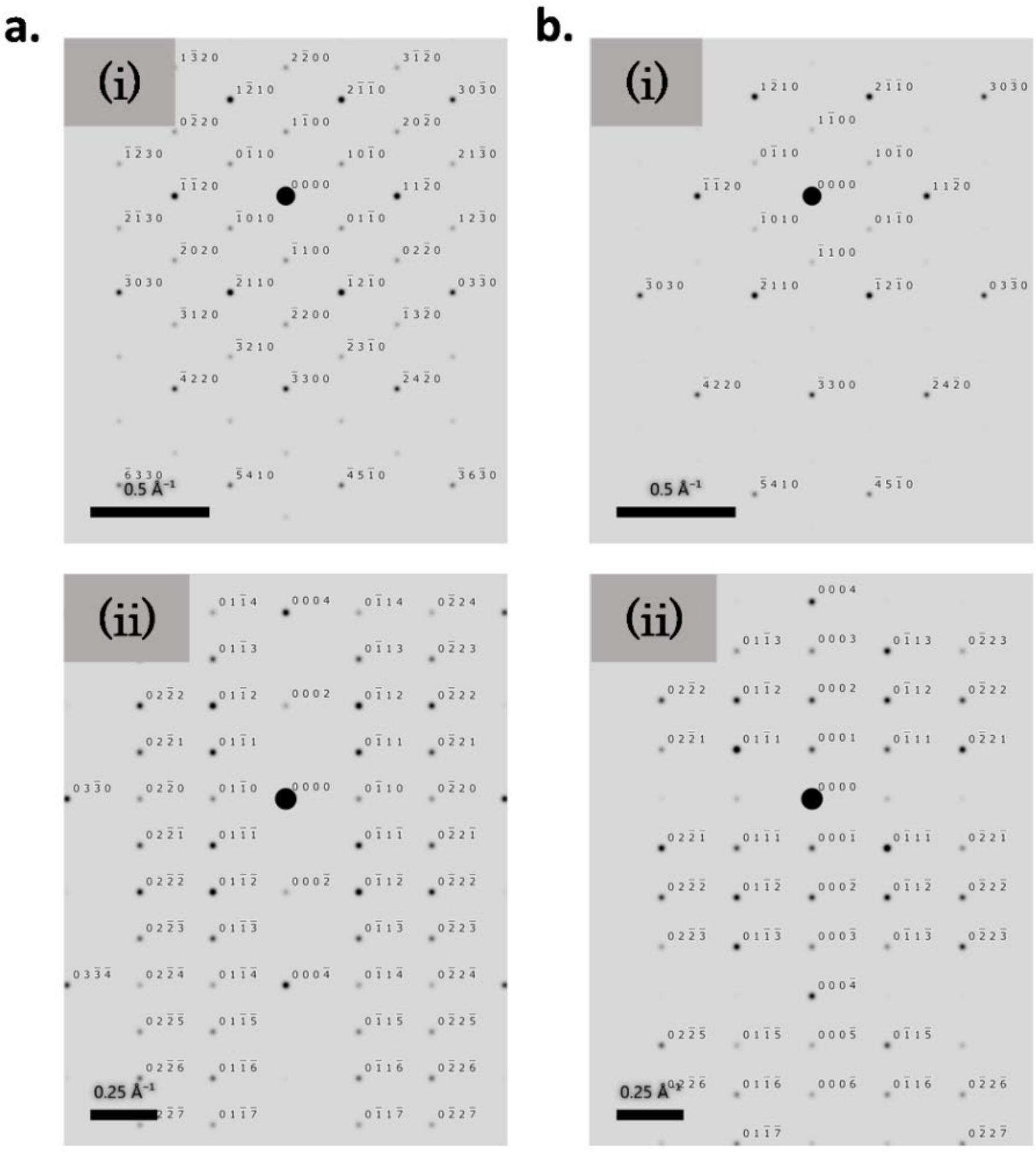

**Figure S7 The simulated electron diffraction pattern of (a) PdTe and (b) $PdTe_2$. The electron beams are parallel to (i) <0001> and (ii) <1010> in (a) and (b).**

## 4. TEM results

The other STEM images obtained in the device II are summarized in the following section.

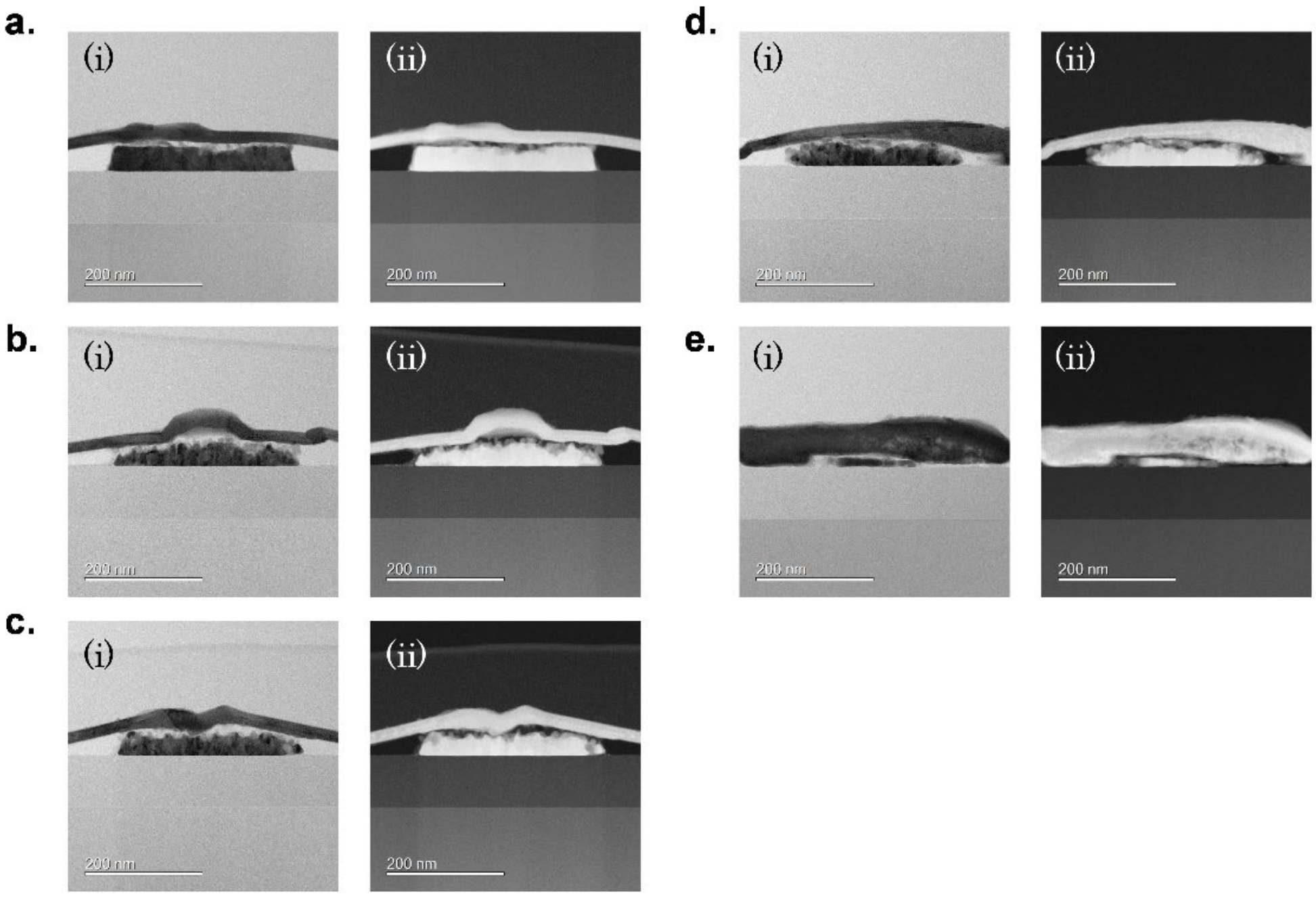

**Figure S8 (a-e) STEM images of electrode A-E, respectively. (i) overall, (ii) BF-STEM, and (iii) HAADF-STEM images are shown, respectively.**

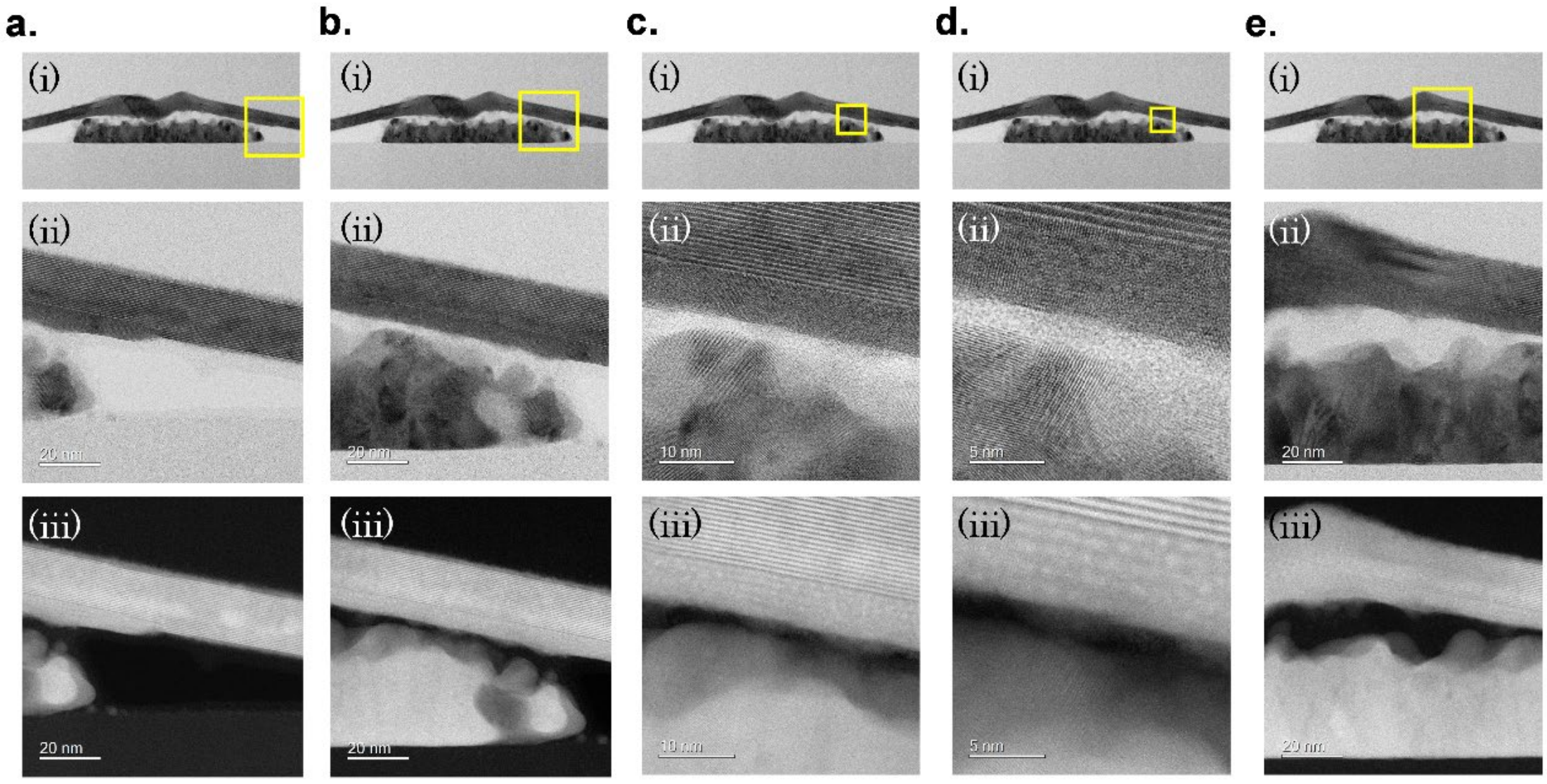

Figure S9 STEM images of electrode C. (i) overall, (ii) BF-STEM, and (iii) HAADF-STEM images are shown, respectively. The area indicated by yellow rectangle are enlarged in (ii) and (iii).

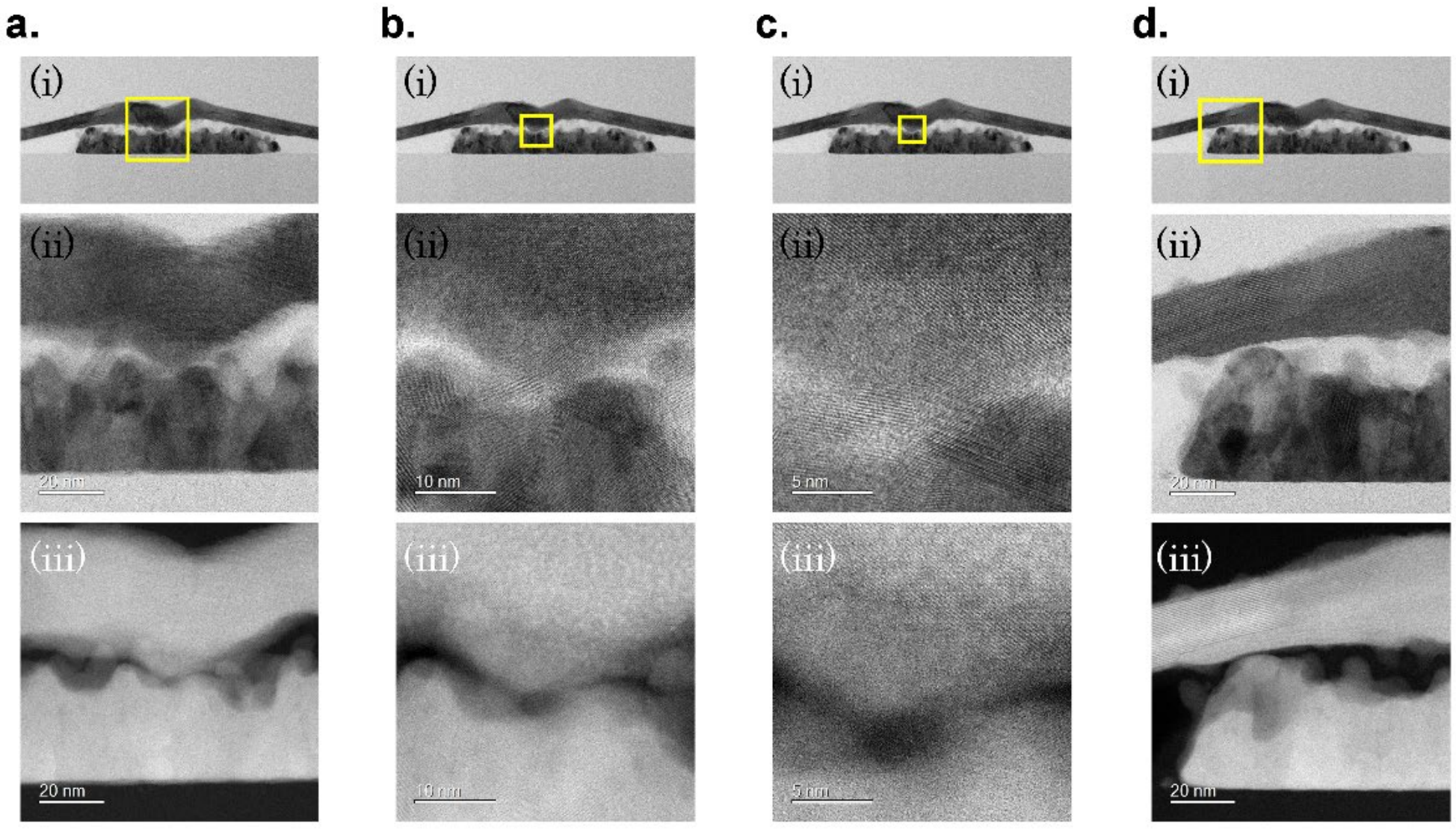

Figure S10 STEM images of electrode C. (i) overall, (ii) BF-STEM, and (iii) HAADF-STEM images are shown, respectively. The area indicated by yellow rectangle are enlarged in (ii) and (iii).

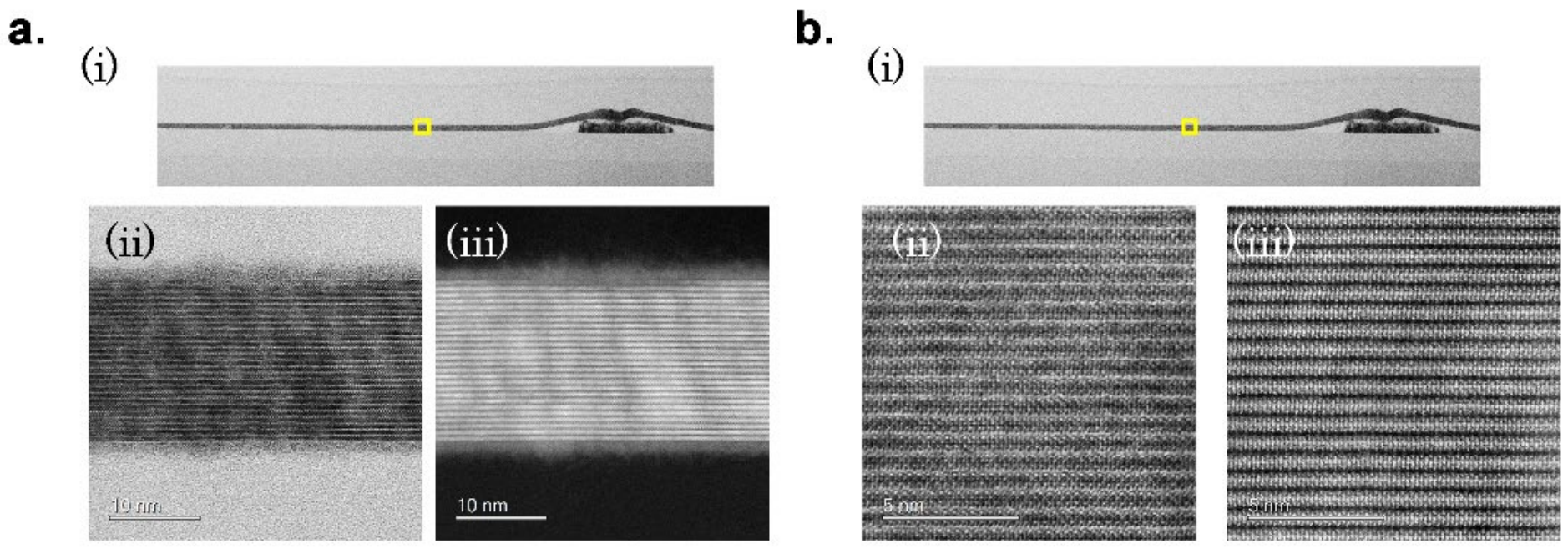

**Figure S11 The STEM images of $WTe_2$ channel between electrode B and C. (i) overall, (ii) BF-STEM, and (iii) HAADF-STEM images are shown, respectively. The area indicated by yellow rectangle are enlarged in (ii) and (iii).**

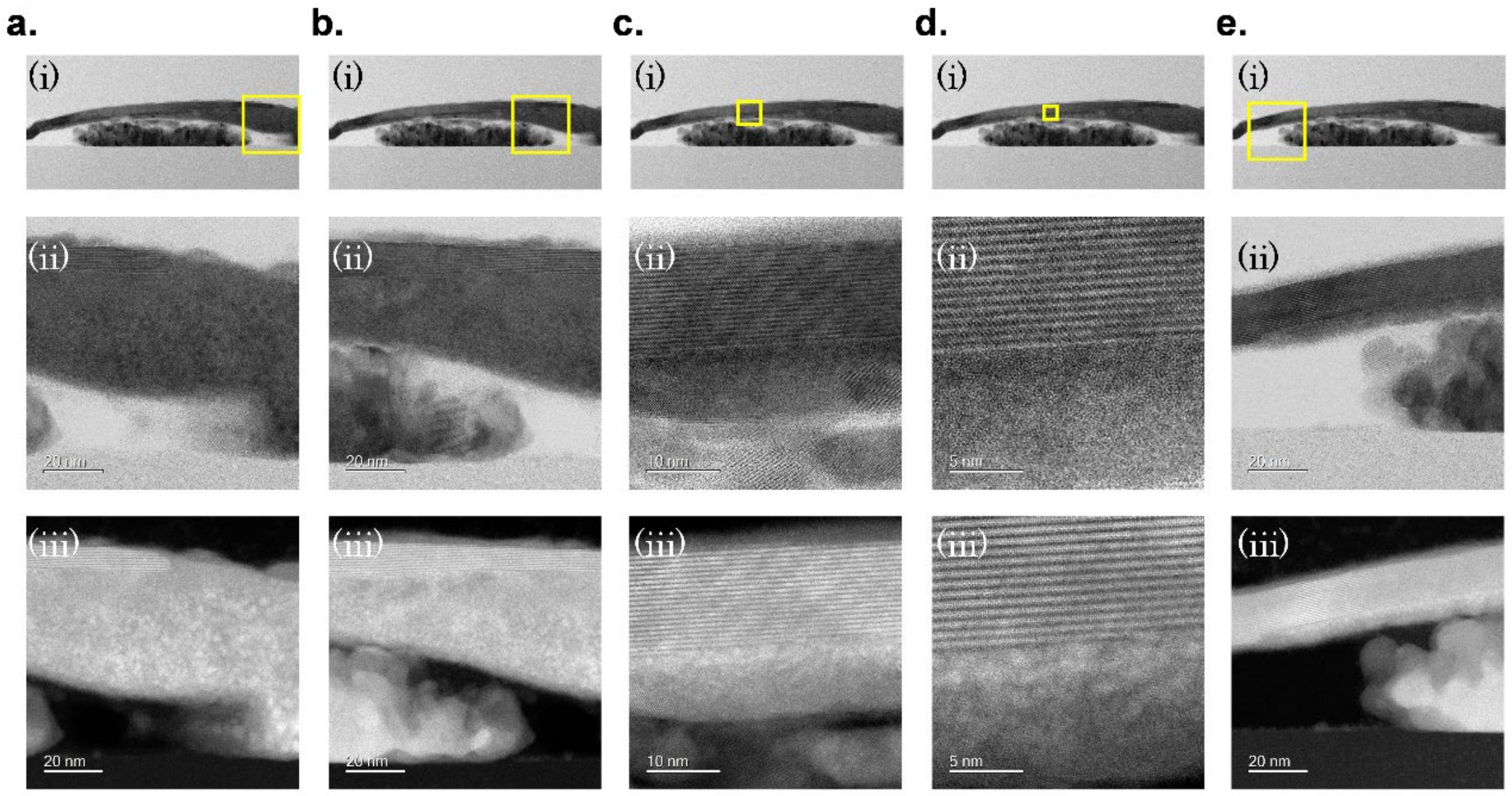

**Figure S12 STEM images of electrode D. (i) overall, (ii) BF-STEM, and (iii) HAADF-STEM images are shown, respectively. The area indicated by yellow rectangle are enlarged in (ii) and (iii).**